# Quantum Entanglement in Dirac Dynamics via Continuous-Time Quantum Walks in a Quantum Circuit Framework

Wei-Ting Wang[1*], Yen-Jui Chang[2,3], Ching Ray Chang[1,2]

[1]Department of Physics, National Taiwan University; Taipei, 106216, Taiwan

[2]Department of Physics and Quantum Information Center, Chung Yuan Christian University; Taoyuan, 320314, Taiwan

[3]Master Program in Intelligent Computing and Big Data, Chung Yuan Christian University, Taoyuan City, Taiwan

*Email:d11245002@ntu.edu.tw

## ABSTRACT

We propose a Continuous-Time Quantum Walks (CTQW) model for one-dimensional Dirac dynamics simulation with higher-order approximation. Our model bridges CTQW with a discrete-time model called Dirac Cellular Automata (DCA) via Quantum Fourier Transformation (QFT). From our continuous-time model, we demonstrate how varying time intervals and position space sizes affect both quantum entanglement between the internal space and external (position) space of the quantum state and the relativistic effect called Zitterbewegung. We find that the time interval changes the transition range for each site, and the position space sizes affect the value of transition amplitude. Therefore, it shows that the size of spacetime plays a crucial role in the observed quantum entanglement and relativistic phenomena in quantum computers. These results enhance the understanding of the interplay between internal and external spaces in Dirac dynamics through the insights of quantum information theory and enrich the application of the quantum walks-based algorithm.

## 1   Introduction

Unlike classical particles, relativistic particles exhibit unique characteristics such as spin, chirality, and the existence of antiparticles [1]. Therefore, simulating the dynamics of relativistic particles is an important topic. Quantum computers are recognized as powerful tools for simulating quantum dynamics, serving as a bridge between quantum physics and quantum information science [2]. To date, there are a lot of quantum circuit models to recover the dynamics of relativistic particles. For Dirac dynamics simulation, discrete-time quantum walks (DTQW) [3-5], and Dirac cellular automata (DCA) [6-8] are more popular. They recover the Dirac dynamics with first-order and second-order approximation respectively. These models reproduce relativistic effects like Zitterbewegung and Klein tunneling on quantum computers [9,10]. In addition, these models are also applied in studying phenomena such as Anderson localization [11-13], topological phases [14,15], and neutrino oscillations [16,17], showcasing their versatility in simulating a wide range of quantum effects. While DTQW and DCA are well-explored for Dirac dynamics, the continuous-time quantum walks (CTQW) [18] model has not been thoroughly investigated. This presents an opportunity to explore delicate phenomena in Dirac dynamics.

We start from Dirac dynamics and then build the CTQW model with the tunable time interval. The corresponding quantum circuit is constructed and runs on a quantum circuit simulator. After that, we demonstrate the Zitterbewegung and the quantum entanglement between the internal space and external space under the influence of time interval of CTQW and the size of the external space.

## 2 Result
### 2.1 The one-dimensional Dirac dynamics

The dynamics of (1+1)-dimensional spin-1/2 particle can be described by the Dirac Hamiltonian:

$$H = -i\gamma^0\gamma^1\partial_x + m\gamma^0, \tag{1}$$

where $\gamma^\mu$ satisfy anti-commutation relation $\gamma^\mu\gamma^\nu + \gamma^\nu\gamma^\mu = 2g^{\mu\nu}I$. This relation makes Lorentz invariant, meaning it respects the principles of special relativity and allows it to describe particles that travel at speeds close to the speed of light.

In momentum space, we have $k = -i\partial_x$ and then rewrite Eq. (1) as

$$H = -\sigma_z k + m\sigma_x. \tag{2}$$

Here, we choose $\gamma^0\gamma^1 = -\sigma_z$ and $\gamma^0 = \sigma_x$. This Dirac Hamiltonian implies that even in one dimension, the Dirac equation describes a spin-½ particle. Specifically, the solution is the wave function, represented as a two-component spinor, where the two components correspond to different states related to spin and/or particle-antiparticle degrees of freedom and predict the existence of antiparticles.

By using Trotter-Suzuki expansion [19], the second-order approximation of the time evolution operator with a small-time interval $\delta t$ is

$$\hat{U}(\delta t) \approx \begin{pmatrix} \hat{Q}_-^{\delta t} & 0 \\ 0 & I \end{pmatrix} \begin{pmatrix} \cos\left(\frac{\theta}{2}\right) & -i\sin\left(\frac{\theta}{2}\right) \\ -i\sin\left(\frac{\theta}{2}\right) & \cos\left(\frac{\theta}{2}\right) \end{pmatrix} \begin{pmatrix} I & 0 \\ 0 & \hat{Q}_+^{\delta t} \end{pmatrix}$$

$$= \begin{pmatrix} \cos\left(\frac{\theta}{2}\right)\hat{Q}_-^{\delta t} & -i\sin\left(\frac{\theta}{2}\right)I \\ -i\sin\left(\frac{\theta}{2}\right)I & \cos\left(\frac{\theta}{2}\right)\hat{Q}_+^{\delta t} \end{pmatrix} \tag{3}$$

where $\hat{Q}_+ = e^{-ik}$, $\hat{Q}_- = \hat{Q}_+^\dagger = e^{ik}$, $\theta = m\delta t$ and $I$ is an identity matrix with the same size as $\hat{Q}_-$ and $\hat{Q}_+$.

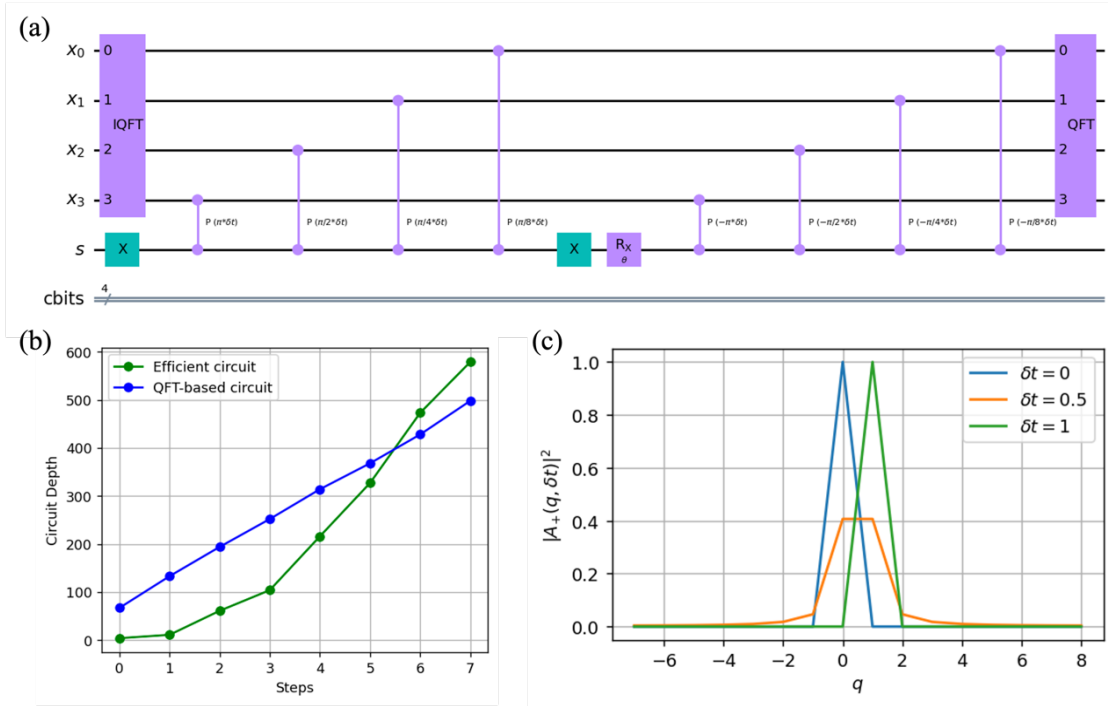

Figure 1 (a) The quantum circuit of continuous-time quantum walks model for Dirac dynamics with the size of position space $N = 16$. (b) The circuit depth of circuit proposed by Ref. [20] (green line) and our method (blue line). (c) The transition amplitude of translation operator $\hat{P}_+(\delta t)$ from a fixed site to its $q$-th neighbor (positive $q$ means right-hand site and negative $q$ means left-hand site).

One of these properties of Dirac time evolution is Zitterbewegung, which is a relativistic quantum effect where a free Dirac particle exhibits rapid oscillatory motion. This arises from the interference between positive and negative energy states (particles and antiparticles). On the other hand, Zitterbewegung, also linked to anomalous velocity, refers to an unexpected velocity component in the motion of a Dirac particle. That is, the velocity operator of the Dirac particle is non-commute with its Hamiltonian.

## 2.2 Simulate Dirac dynamics in quantum computers

Considering this quantum system lives on a finite Hilbert space with $N$ sites, we have quantized toroidal phase space, in other words, $-\pi < k \leq \pi$. It means that we have a discrete spacetime with time interval $\delta t$ and space interval $\delta x = 1$, and the state of Dirac particle is described by quantum field $\Psi(x)$, where $x \in \mathbb{Z}$. The corresponding quantum circuit of this time evolution operator is demonstrated in Figure 1(a). It has a similar circuit depth to the approach proposed in Ref. [20,21] without limitation (see Figure 1(b)). The quantum Fourier transform (QFT) operator $\hat{F}$ maps the quantum state from real space to momentum space with matrix representation:

$$\hat{F} = \begin{pmatrix} 1 & 1 & 1 & \cdots & 1 \\ 1 & \omega & \omega^2 & \cdots & \omega^{N-1} \\ 1 & \omega^2 & \omega^4 & \cdots & \omega^{2(N-1)} \\ \vdots & \vdots & \vdots & \ddots & \vdots \\ 1 & \omega^{N-1} & \omega^{2(N-1)} & \cdots & \omega^{(N-1)(N-1)} \end{pmatrix}_{N \times N}. \quad (4)$$

Thus, the $\hat{Q}_+$ has a matrix representation

$$\hat{Q}_+ = \begin{pmatrix} 1 & 0 & 0 & \cdots & 0 \\ 0 & \omega & 0 & \ddots & 0 \\ 0 & 0 & \omega^2 & \ddots & \vdots \\ \vdots & \ddots & \ddots & \ddots & 0 \\ 0 & \cdots & 0 & 0 & \omega^{(N-1)} \end{pmatrix}_{N \times N} \quad (5)$$

and satisfies $\hat{Q}_+^N = \hat{Q}_-^N = I$, where $\omega = e^{2\pi i/N}$.

On the other hand, when we want to analyze the interaction between sites, we go back to the real state and have the translation operator $\hat{P}_+(\delta t)$, that is

$$\hat{P}_+(\delta t) = \hat{F}\hat{Q}_+^{\delta t}\hat{F}^\dagger$$

$$= \frac{1}{N}\sum_{j=0}^{N-1} \begin{pmatrix} \omega^{j\delta t} & \omega^{j\delta t}\omega^{-j} & \omega^{j\delta t}\omega^{-2j} & \cdots & \omega^{j\delta t}\omega^{j} \\ \omega^{j\delta t}\omega^{j} & \omega^{j\delta t} & \omega^{j\delta t}\omega^{-j} & \cdots & \omega^{j\delta t}\omega^{2j} \\ \omega^{j\delta t}\omega^{2j} & \omega^{j\delta t}\omega^{j} & \omega^{j\delta t} & \cdots & \omega^{j\delta t}\omega^{3j} \\ \vdots & \vdots & \vdots & \ddots & \vdots \\ \omega^{j\delta t}\omega^{-j} & \omega^{j\delta t}\omega^{-2j} & \omega^{j\delta t}\omega^{-3j} & \cdots & \omega^{j\delta t} \end{pmatrix} \quad (6)$$

Therefore, the transition amplitude from any arbitrary site to its $q$-th neighbor is

$$A_q^{(+)}(\delta t) = \frac{1}{N}\sum_{j=0}^{N-1} \omega^{j(\delta t - q)} = \frac{1}{N}\frac{1 - \omega^{N(\delta t - q)}}{1 - \omega^{(\delta t - q)}} = \frac{1}{N}\frac{1 - e^{i2\pi(\delta t - q)}}{1 - e^{\frac{i2\pi(\delta t - q)}{N}}}, \quad (7)$$

where positive $q$ means right-hand site and negative $q$ means left-hand site.
For the infinite space, we have

$$\left|A_q^{(+)}(\delta t)\right|^2 = \frac{1}{N^2}\frac{1 - \cos(2\pi(\delta t - q))}{1 - \cos(\frac{2\pi}{N}(\delta t - q))}$$

$$= \frac{1}{N^2}\frac{1 - \cos(2\pi(\delta t - q))}{\frac{1}{2}\left(\frac{2\pi}{N}(\delta t - q)\right)^2}$$

$$= \frac{1 - \cos(2\pi(\delta t - q))}{2\pi^2(\delta t - q)^2} \quad (8)$$

with $N \to \infty$.

Summarily, for $\hat{Q}_-$, the transition amplitude from any arbitrary site to its $q$-th neighbor is

$$A_q^{(-)}(\delta t) = \frac{1}{N}\sum_{j=0}^{N-1} \omega^{j(\delta t + q)} = \frac{1}{N}\frac{1 - \omega^{N(\delta t + q)}}{1 - \omega^{(\delta t + q)}} = \frac{1}{N}\frac{1 - e^{i2\pi(\delta t + q)}}{1 - e^{\frac{i2\pi(\delta t + q)}{N}}} \quad (9)$$

where positive $q$ means right-hand site and negative $q$ means left-hand site. And when $N \to \infty$, we have

$$\left|A_q^{(+)}(\delta t)\right|^2 = \frac{1 - \cos(2\pi(\delta t + q))}{2\pi^2(\delta t + q)^2} \qquad (10)$$

Therefore, from Eq. (2), the state in time $t + \delta t$ is described by

$$\begin{pmatrix} \psi_R(x, t + \delta t) \\ \psi_L(x, t + \delta t) \end{pmatrix} = \cos\left(\frac{\theta}{2}\right) \sum_{j=0}^{N-1} \begin{pmatrix} A_q^{(-)}(\delta t)\psi_R(x + q, t) \\ A_q^{(+)}(\delta t)\psi_L(x - q, t) \end{pmatrix} - i\sin\left(\frac{\theta}{2}\right)\begin{pmatrix} \psi_L(x, t) \\ \psi_R(x, t) \end{pmatrix}. \qquad (11)$$

In addition, the $\delta t$ determines the interaction range, and the particle has a wider interaction range while $0 < \delta t < 1$ (see Figure 1(c)). When $\delta t = 1$, we obtain Dirac cellular automata with the nearest neighborhood interaction, that is

$$\hat{U}(\delta t = 1) = \begin{pmatrix} \cos\left(\frac{\theta}{2}\right)\hat{P}_+ & -i\sin\left(\frac{\theta}{2}\right)I \\ -i\sin\left(\frac{\theta}{2}\right)I & \cos\left(\frac{\theta}{2}\right)\hat{P}_- \end{pmatrix} \qquad (12)$$

and

$$\begin{pmatrix} \psi_R(x, t + 1) \\ \psi_L(x, t + 1) \end{pmatrix} = \begin{pmatrix} \cos\left(\frac{\theta}{2}\right)\psi_R(x + 1, t) - i\sin\left(\frac{\theta}{2}\right)\psi_L(x, t) \\ \cos\left(\frac{\theta}{2}\right)\psi_L(x - 1, t) - i\sin\left(\frac{\theta}{2}\right)\psi_R(x, t) \end{pmatrix}, \qquad (13)$$

where

$$\hat{P}_+ = \hat{P}_+(\delta t = 1) = \hat{P}_+ = \begin{pmatrix} 0 & 1 & 0 & \cdots & 0 \\ 0 & 0 & 1 & \ddots & \vdots \\ \vdots & 0 & 0 & \ddots & 0 \\ 0 & \ddots & \ddots & \ddots & 1 \\ 1 & 0 & \cdots & 0 & 0 \end{pmatrix}_{N \times N} \qquad (14)$$

and

$$\hat{P}_+ = \hat{P}_+(\delta t = 1) = \hat{P}_+ = \begin{pmatrix} 0 & 0 & 0 & \cdots & 1 \\ 1 & 0 & 0 & \ddots & 0 \\ 0 & 1 & 0 & \ddots & 0 \\ \vdots & \ddots & \ddots & \ddots & \vdots \\ 0 & 0 & \cdots & 1 & 0 \end{pmatrix}_{N \times N} \qquad (15)$$

### 2.3 Quantum entanglement and Zitterbewegung

The internal degrees of freedom of a Dirac particle are related to its spinor structure called internal space $\mathcal{H}_c$. The external degrees of freedom of a Dirac particle are associated with its position and momentum in space called external space $\mathcal{H}_p$. The Dirac dynamics couples the internal and external space interestingly. From the perspective of quantum information theory, the internal space can become entangled with the external space. We simulate the quantum entanglement and Zitterbewegung and the relative between the initial state, mass, and time interval. Here we quantify the strength of entanglement at time $t$ by the von Neumann entropy:

$$S(\rho_c(t)) = -\text{Tr}_c(\rho_c(t)\log(\rho_c(t))) = S(\rho_x(t)), \qquad (16)$$

where $\rho_c(t) = \text{Tr}_x(\rho(t))$ and $\rho_x(t) = \text{Tr}_c(\rho(t))$ are reduced density matrices at time $t$ which take partial trace respect to external space $\mathcal{H}_p$ and internal space $\mathcal{H}_c$, respectively. From our previous study [22], for the Zitterbewegung effect, we examine the expectation value of velocity $-\langle Z \rangle$ on the internal space. As demonstrated in Fig. 2, we plot the time evolution spacetime diagram and its corresponding velocity and entropy. The higher quantum entanglement strength induces the oscillation of velocity. The initial state is a highly localized wave function fixed in the $x = 0$. In Figure 2(a), the massless particle has no Zitterbewegung oscillation and entanglement. However, in Figure 2(b), when the massive particle is considered, the entanglement has been created and has the same oscillation frequency.

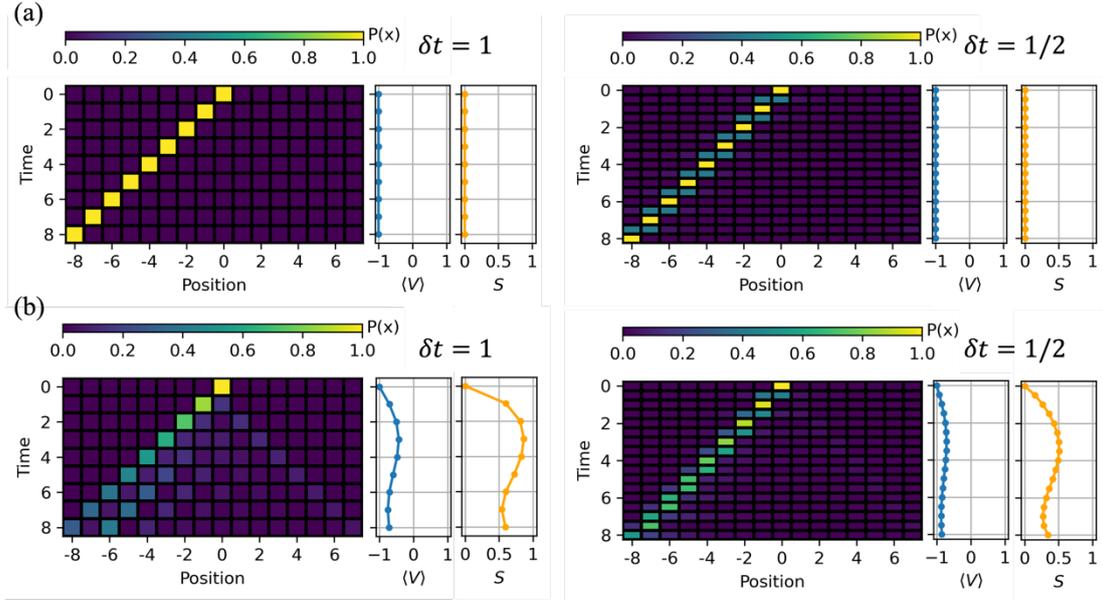

Figure 2 The spacetime diagram of the Dirac particle produced by the continuous-time model with initial state $\Psi(t=0) = |0\rangle \otimes |x=0\rangle$. The time interval is $\delta t = 1$ for the left-hand side, and the time interval is $\delta t = 1/2$ for the right-hand side. The blue line represents the expectation value of velocity $\langle V \rangle$ and the orange line represents the quantum entanglement between the internal space and external space. (a) The particle is massless, and no entanglement or Zitterbewegung oscillations are observed. (b) The particle is massive, and we observe entanglement or Zitterbewegung oscillations.

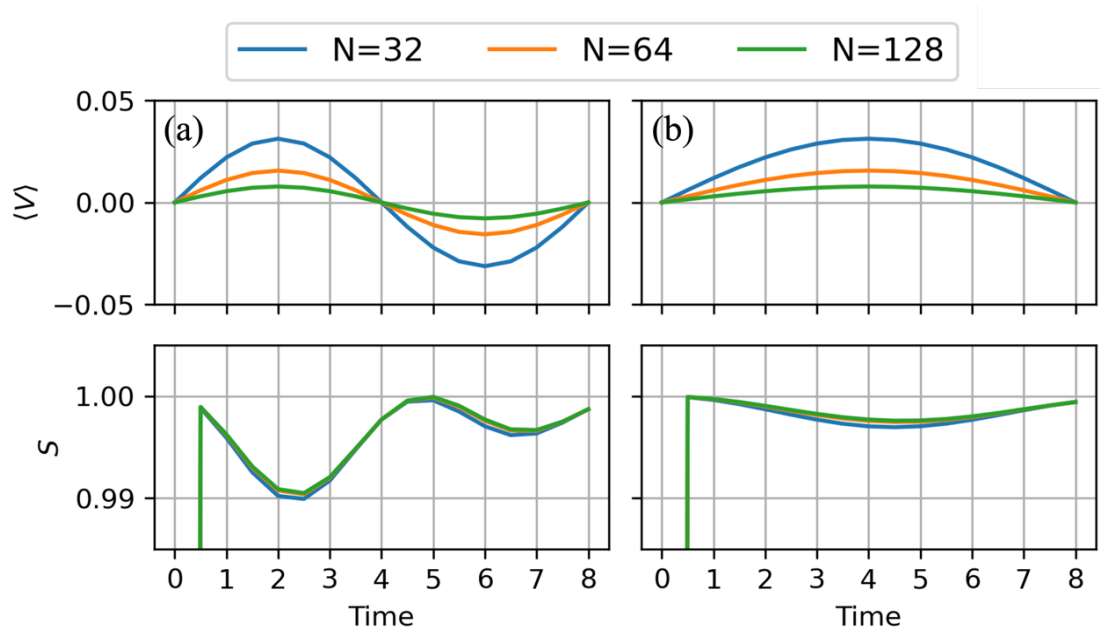

Figure 3 The quantum entanglement between the internal space and the expectation value of velocity under external space size $N = 32, 64$, and $128$. (a) The mass of the particle is $m = \pi/4$. (b) The mass of the particle is $m = \pi/8$.

Moreover, as mentioned in section 2.2, the interaction between each site correlated with the size of the external space $\mathcal{H}_p$, affecting the entanglement and expectation value of velocity. As shown in Figure 3, three different sizes are considered. The result shows that the particle's mass is related to the Zitterbewegung frequency, and with bigger internal space, we obtained higher entanglement and smaller Amplitude of Zitterbewegung.

## 3 Conclusion

Our study explores the dynamics of a Dirac particle in a CTQW model via QFT, which is proven to be an efficient algorithm to implement in quantum computers [23]. Then we focus on how quantum entanglement and relativistic effects such as Zitterbewegung oscillation are influenced by key parameters like the particle's mass and the time interval. When the time interval equals the space, the system is the discrete-time model called DCA, where the particle only interacts with its nearest neighbors. By simulating the interaction between the internal and external degrees of freedom of the Dirac particle, we reveal several important findings. First, the coupling between the internal and external spaces generates quantum entanglement, influencing the Amplitude of Zitterbewegung. For massless particles, no entanglement or Zitterbewegung oscillations are observed. On the other hand, for massive particles, both effects emerge, and the Zitterbewegung frequency is shown to be directly related to the particle's mass and the initial state. Secondly, except the mass and the initial state, we also find that the impact of external space size and time interval affects the interaction range and the strength of entanglement.

This study demonstrates that the size of spacetime plays a crucial role in the observed quantum entanglement and relativistic phenomena in quantum computers. These results enhance the understanding of the interplay between internal and external spaces in

Dirac dynamics, offering insights into how quantum information theory can shed light on relativistic quantum systems.

**References**

1. Pal, P. B. Dirac, majorana, and weyl fermions. *Am. J. Phys.* **79**, 485-498 (2011)
2. Fillion-Gourdeau, F., MacLean, S. & Laflamme, R. Algorithm for the solution of the Dirac equation on digital quantum computers. Phys. Rev. A 95, 042343 (2017).
3. Strauch, F. W. Relativistic quantum walks. *Phys. Rev. A* **73**, 054302 (2006).
4. Bracken, A., Ellinas, D. & Smyrnakis, I. Free-Dirac-particle evolution as a quantum random walk. *Phys. Rev. A* **75**, 022322 (2007).
5. Chandrashekar, C., Banerjee, S. & Srikanth, R. Relationship between quantum walks and relativistic quantum mechanics. *Phys. Rev. A* **81**, 062340 (2010).
6. Bialynicki-Birula, I. Weyl, Dirac, and Maxwell equations on a lattice as unitary cellular automata. *Phys. Rev. D* **49**, 6920 (1994).
7. Meyer, D. A. From quantum cellular automata to quantum lattice gases. *J. Stat. Phys.* **85**, 551-574 (1996).
8. Bisio, A., D'Ariano, G. M. & Tosini, A. Quantum field as a quantum cellular automaton: The Dirac free evolution in one dimension. *Ann. Phys.* **354**, 244-264 (2015).
9. Kurzyński, P. Relativistic effects in quantum walks: Klein's paradox and zitterbewegung. *Phys. Lett. A* **372**, 6125-6129 (2008).
10. Bisio, A., D'Ariano, G. M. & Tosini, A. Dirac quantum cellular automaton in one dimension: Zitterbewegung and scattering from potential. *Phys. Rev. A* **88**, 032301 (2013).
11. Schreiber, A. *et al.* Decoherence and disorder in quantum walks: from ballistic spread to localization. *Phys. Rev. Lett.* **106**, 180403 (2011).
12. Crespi, A. *et al.* Anderson localization of entangled photons in an integrated quantum walk. *Nat. Photonics* **7**, 322-328 (2013).
13. Derevyanko, S. Anderson localization of a one-dimensional quantum walker. *Sci. Rep.* **8**, 1-11 (2018).
14. Kitagawa, T., Rudner, M. S., Berg, E. & Demler, E. Exploring topological phases with quantum walks. *Phys. Rev. A* **82**, 033429 (2010).
15. Flurin, E. *et al.* Observing topological invariants using quantum walks in superconducting circuits. *Phys. Rev. X* **7**, 031023 (2017).
16. Di Molfetta, G. & Pérez, A. Quantum walks as simulators of neutrino oscillations in a vacuum and matter. *New J. Phys.* **18**, 103038 (2016).
17. Mallick, A., Mandal, S. & Chandrashekar, C. Neutrino oscillations in discrete-time quantum walk framework. *Eur. Phys. J. C* **77**, 1-11 (2017).

18. Childs, A. M., Farhi, E. & Gutmann, S. An example of the difference between quantum and classical random walks. *Quantum Inf. Process.* **1**, 35-43 (2002).
19. Berry, D.W., Ahokas, G., Cleve, R. *et al.* Efficient Quantum Algorithms for Simulating Sparse Hamiltonians. *Commun. Math. Phys.* **270**, 359–371 (2007).
20. Huerta Alderete, C. et al. Quantum walks and Dirac cellular automata on a programmable trapped-ion quantum computer. Nat. Commun. 11, 1-7 (2020).
21. Singh, S. et al. Quantum circuits for the realization of equivalent forms of one-dimensional discrete-time quantum walks on near-term quantum hardware. Phys. Rev. A 104, 062401 (2021).
22. Wang, W. T., He, X. G., Kao, H. C., & Chang, C. R. (2024). Observing Majorana Fermion dynamic properties on a NISQ computer. Chinese Journal of Physics. 90. 10.1016/j.cjph.2024.05.006.
23. Shakeel, A. Efficient and scalable quantum walk algorithms via the quantum Fourier transform. Quantum Inf. Process. 19, 1-26 (2020).